\documentclass[9pt,conference]{IEEEtran}

\usepackage[preprint]{waspaa25}

\usepackage{bm} 

\title{Real-Time System for Audio-Visual Target Speech Enhancement}

\name{Teng Ma$^{1,2}$,
      Sile Yin$^{1}$,
      Li-Chia Yang$^{1}$,
      Shuo Zhang$^{1}$}
\address{$^{1}$Bose Corporation, Framingham, USA \;
$^{2}$Georgia Institute of Technology, Atlanta, USA\\
}

\begin{document}

\maketitle

\begin{abstract}
We present a live demonstration for RAVEN, a \textbf{r}eal-time \textbf{a}udio-\textbf{v}isual speech \textbf{en}hancement system designed to run entirely on a CPU. In single-channel, audio-only settings, speech enhancement is traditionally approached as the task of extracting clean speech from environmental noise. More recent work has explored the use of visual cues, such as lip movements, to improve robustness, particularly in the presence of interfering speakers. However, to our knowledge, no prior work has demonstrated an interactive system for real-time audio-visual speech enhancement operating on CPU hardware.
RAVEN fills this gap by using pretrained visual embeddings from an audio-visual speech recognition model to encode lip movement information. The system generalizes across environmental noise, interfering speakers, transient sounds, and even singing voices. In this demonstration, attendees will be able to experience live audio-visual target speech enhancement using a microphone and webcam setup, with clean speech playback through headphones. 
\end{abstract}

\section{Introduction}
\label{sec:intro}

Speech enhancement is commonly defined as the removal of background noise \cite{das_fundamentals_2021}, often focused on environmental sounds. However, real-world acoustic environments are more complex, frequently involving transient noises such as sudden bangs, as well as interfering speakers. Traditional audio-only speech enhancement algorithms struggle with separating out clean speech from competing speakers, unless provided with enrollment audio from target speaker. These challenges prompt researchers to explore the use of additional modalities, such as visual information, to improve target speech enhancement performance in recent years, as computing power increases\cite{adeel_lip-reading_2021}. There has been a rise in audio-visual speech enhancement (AVSE) literature, from mask-based approaches \cite{afouras2018conversation, wang_2020_mask, ephrat_looking_2018} to synthesis-based approaches\cite{Yang_2022_CVPR, sadeghi_2020_VAE, jung_flowavse_2024}, achieving better results than only using audio\cite{mira_2023_lavoce, chen_rt--voce_2024}. 

\subsection{Motivation}

\begin{figure}[h]
  \centering
  \centerline{\includegraphics[width=\columnwidth]{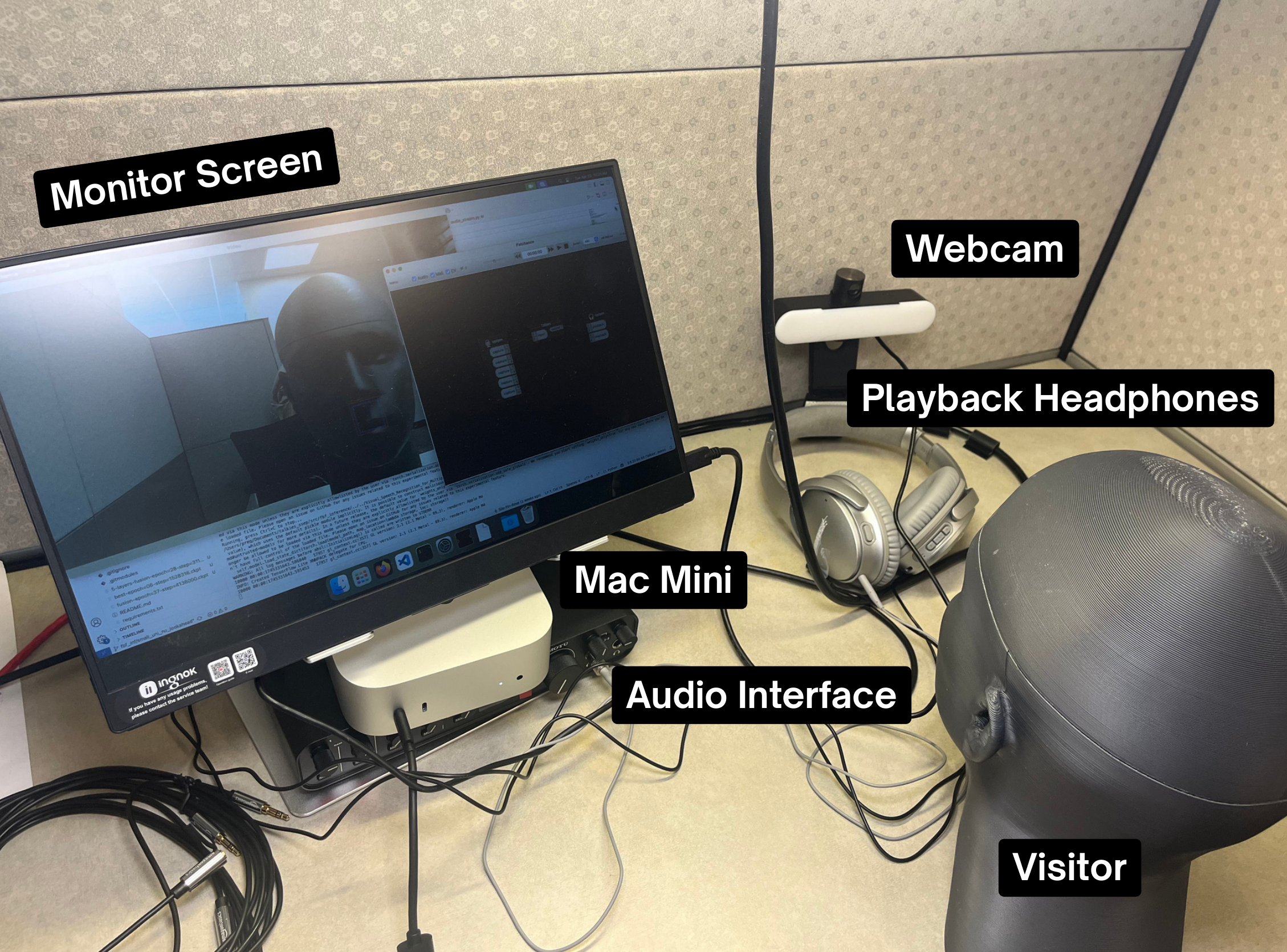}}
  \caption{\textbf{Setup of our real-time audio-visual speech enhancement system.} The audio-visual input is streamed into our Python application via the webcam and the microphones. Visitors can hear the clean speech through the playback headphones in real time. }
  \label{fig:system}
\end{figure}

Despite this increase in AVSE research, most existing systems operate in a non-causal setting\cite{ephrat_looking_2018, mira_2023_lavoce, afouras2018conversation}. While having access to future information may achieve strong performance, they are unsuitable for real-time deployment. Although some real-time AVSE systems have been proposed, their implementations and demonstrations are not made available, making reproduction difficult \cite{gogate_cochleanet_2020, chen_rt--voce_2024, zhu_real-time_2023}. Motivated by this gap, we present a real-time AVSE algorithm and built an application that enables users to test the speech enhancement algorithm in realistic scenarios, as seen in \cref{fig:system}. To our knowledge, this is the first publicly available, real-time AVSE demonstration that can be run on a CPU.

\subsection{Application \& Problem Scenario}

Real-time AVSE systems are especially useful for video calls\cite{inan19_interspeech}, where the model can isolate and enhance the on-screen speaker’s voice. This also applies to in-car communication systems\cite{chuang_improved_2022}, enabling clear speech pickup even in noisy cabins with multiple speakers or background music.

Beyond video calls, the technology extends to wearables equipped with cameras, such as smart glasses or headphones, where it can provide users with an audio feed of the clean speech of the speaker in their field of view. This offers a potential solution to the longstanding cocktail party problem, which refers to the human ability to focus auditory attention on a single speaker in a noisy, multi-speaker setting\cite{li2018effects}. In addition, AVSE algorithms also hold promise for application in advanced audio-visual hearing aids\cite{adeel2020novel, hou2018audio}. Conventional hearing aids typically amplify all incoming sounds indiscriminately, which can overwhelm users in noisy environments and limit intelligibility. In contrast, audio-visual hearing aids could identify and prioritize speech signal of interest by leveraging relevant visual cues.

\section{System Design}
\label{sec:double_blind}

Our real-time AVSE system consists of two main components: (1) a causal AVSE model designed for real-time inference trained on VoxCeleb2 \cite{Chung2018VoxCeleb2DS}, and (2) a real-time streaming pipeline that interfaces with live audio-visual inputs, processes them through the AVSE model, and outputs the enhanced speech signal. \footnote{The algorithm and training code are released through Interspeech 2025\cite{ma25c_interspeech}; however, the real-time system has never been demonstrated live.}

\subsection{Model Architecture}
\begin{figure*}[t]
  \centering
  \centerline{\includegraphics[width=\textwidth]{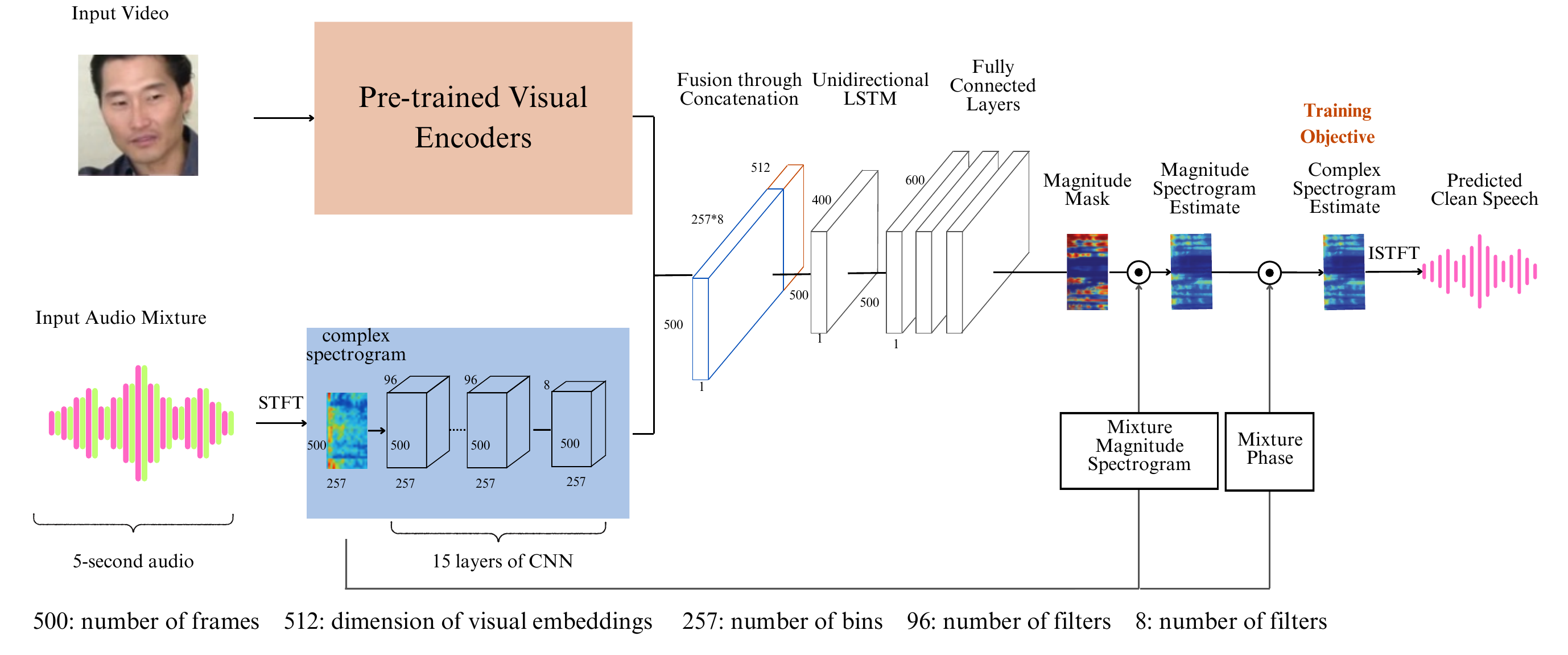}}
  \caption{\textbf{Architecture of our mask-based late fusion method.} A batch normalization layer is added after each CNN layer, and a ReLU layer is added to each CNN and fully connected layer. A Sigmoid function is applied to the magnitude mask to limit the mask values from 0 to 1.}
  \label{fig:architecture}
\end{figure*}

As shown in \cref{fig:architecture}, RAVEN uses a mask-based, late-fusion architecture. The visual stream first crops the mouth region from the input video and passes it through a pretrained visual encoder to extract lip movement embeddings. The pretrained visual encoder, Visual Speech Recognition in the Wild (VSRiW)\cite{ma_end--end_2021, ma_visual_2022}, is an end-to-end visual speech recognition model that integrates feature extraction with a hybrid CTC/attention back-end. The checkpoint that we used was trained on GRID, which results in a Word Error Rate of 4.8.  Its visual front end uses a ResNet encoder with a 3D convolutional layer that has a kernel size of 5. It has a padding of 2 on both sides along the time axis, which results in a receptive field of 5 including a look-ahead of 2 video frames. The visual stream has a frame rate of 25 frames per second. 

The audio input mixture clips are synthesized to simulate our target task by combining the target audio with the audio from another randomly selected input video. Then the signal-to-noise ratio of the input audio mixture is constrained to a range of -5 dB to 5 dB. We obtain the time-frequency representations of these 5-second clips through Short-Time Fourier Transform (STFT). We use a Hann
window of length 400, a hop size of 160, and 512 frequency bins
(nfft), with a power-law compression rate (p) of 0.3. These spectrograms are passed into a 15-layer convolutional neural network (CNN). 

To align the audio and visual streams, the visual embeddings are upsampled to 100 fps before they are concatenated with the audio embeddings. The concatenated audio-visual features are subsequently fed into a uni-directional Long Short-Term Memory (LSTM) network, the output of which is passed through a series of three fully connected layers that produce a predicted magnitude mask. This mask is applied element-wise to the magnitude spectrogram of the noisy input mixture, yielding an estimated magnitude spectrogram of the clean speech. To reconstruct the enhanced speech waveform, the estimated clean magnitude spectrogram is combined with the phase of the original noisy mixture to form an estimated complex spectrogram. The resulting complex spectrogram is then transformed back into the time domain using the inverse Short-Time Fourier Transform (ISTFT).

The model is trained using a phase-sensitive approximation (PSA) loss function, which minimizes the Mean Squared Error (MSE) between both the predicted and ground truth magnitude as well as complex spectrogram to mitigate for the loss of phase information during training, inspired by \cite{erdogan_phase-sensitive_2015, wang_compensation_2021}.

\subsection{Components \& Implementation}

\begin{figure}[h]
  \centering
  \centerline{\includegraphics[width=\linewidth]{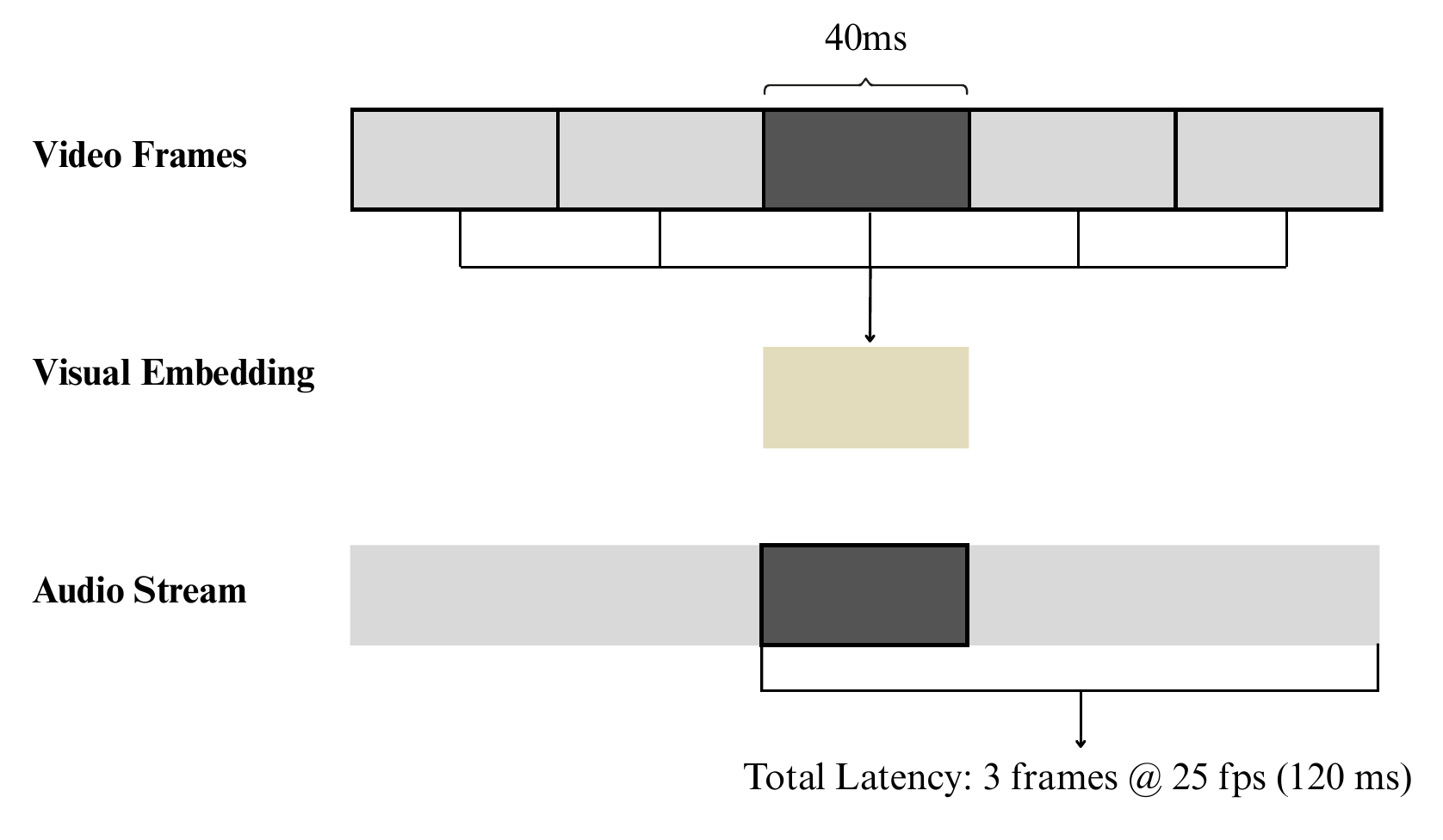}}
  \caption{\textbf{Latency diagram of the RAVEN system.} To generate a meaningful visual embedding for each frame, we buffer 5 frames of audio and video input, streamed at 25 frames per second. This is required by the pretrained visual encoder, which has a receptive field of 5 frames including a 2-frame lookahead. The resulting algorithmic latency is 120 ms. }
  \label{fig:latency}
\end{figure}

The system is implemented in Python. Audio streaming is handled using PyAudio, while video is captured using OpenCV (cv2). Both modalities run at 25 frames per second, corresponding to a 40 ms frame interval. The visual encoder has a receptive field of 5, which includes a 2-frame lookahead. In order to produce a meaningful visual embedding for the current frame, we maintain a buffer of 5 frames. In this setup, the current frame is the middle frame (the third frame) in the buffer. This results in an algorithmic latency of 120 ms, as shown in \cref{fig:latency}. Since the processing time per frame remains below 40 ms, the system meets real-time performance requirements.

\section{Demonstration \& Interaction}

The demonstration runs on a Mac Mini using our Python-based application. Audio-visual input is captured through a microphone and webcam, and the enhanced speech is played back in real time through headphones. As shown in \cref{fig:system}, the setup includes a Mac Mini, a monitor screen, audio interface, a webcam, microphones, and headphones placed on a table.

Visitors will be able to interact with the system in real time. The system works as such: the person that the webcam points at is the target speaker, which could be the visitor themselves or another visitor, and the clean target speech will be played back through the headphone in real time. Visitors will be encouraged to test out different interference conditions, such as clapping or singing, to evaluate the model's robustness and responsiveness.

\clearpage
\bibliographystyle{IEEEtran}
\bibliography{main}

\begin{thebibliography}{10}
\providecommand{\url}[1]{#1}
\csname url@samestyle\endcsname
\providecommand{\newblock}{\relax}
\providecommand{\bibinfo}[2]{#2}
\providecommand{\BIBentrySTDinterwordspacing}{\spaceskip=0pt\relax}
\providecommand{\BIBentryALTinterwordstretchfactor}{4}
\providecommand{\BIBentryALTinterwordspacing}{\spaceskip=\fontdimen2\font plus
\BIBentryALTinterwordstretchfactor\fontdimen3\font minus \fontdimen4\font\relax}
\providecommand{\BIBforeignlanguage}[2]{{%
\expandafter\ifx\csname l@#1\endcsname\relax
\typeout{** WARNING: IEEEtran.bst: No hyphenation pattern has been}%
\typeout{** loaded for the language `#1'. Using the pattern for}%
\typeout{** the default language instead.}%
\else
\language=\csname l@#1\endcsname
\fi
#2}}
\providecommand{\BIBdecl}{\relax}
\BIBdecl

\bibitem{das_fundamentals_2021}
N.~Das, S.~Chakraborty, J.~Chaki, N.~Padhy, and N.~Dey, ``Fundamentals, present and future perspectives of speech enhancement,'' \emph{International Journal of Speech Technology}, vol.~24, no.~4, pp. 883--901, 2021.

\bibitem{adeel_lip-reading_2021}
A.~Adeel, M.~Gogate, A.~Hussain, and W.~M. Whitmer, ``Lip-{Reading} {Driven} {Deep} {Learning} {Approach} for {Speech} {Enhancement},'' \emph{IEEE Transactions on Emerging Topics in Computational Intelligence}, vol.~5, no.~3, pp. 481--490, Jun. 2021.

\bibitem{afouras2018conversation}
T.~Afouras, J.~S. Chung, and A.~Zisserman, ``The conversation: Deep audio-visual speech enhancement,'' \emph{arXiv preprint arXiv:1804.04121}, 2018.

\bibitem{wang_2020_mask}
W.~Wang, C.~Xing, D.~Wang, X.~Chen, and F.~Sun, ``A robust audio-visual speech enhancement model,'' in \emph{ICASSP 2020 - 2020 IEEE International Conference on Acoustics, Speech and Signal Processing (ICASSP)}, 2020, pp. 7529--7533.

\bibitem{ephrat_looking_2018}
A.~Ephrat, I.~Mosseri, O.~Lang, T.~Dekel, K.~Wilson, A.~Hassidim, W.~T. Freeman, and M.~Rubinstein, ``Looking to {Listen} at the {Cocktail} {Party}: {A} {Speaker}-{Independent} {Audio}-{Visual} {Model} for {Speech} {Separation},'' \emph{ACM Transactions on Graphics}, vol.~37, no.~4, pp. 1--11, 2018.

\bibitem{Yang_2022_CVPR}
K.~Yang, D.~Markovi\'c, S.~Krenn, V.~Agrawal, and A.~Richard, ``Audio-visual speech codecs: Rethinking audio-visual speech enhancement by re-synthesis,'' in \emph{Proceedings of the IEEE/CVF Conference on Computer Vision and Pattern Recognition (CVPR)}, June 2022, pp. 8227--8237.

\bibitem{sadeghi_2020_VAE}
M.~Sadeghi, S.~Leglaive, X.~Alameda-Pineda, L.~Girin, and R.~Horaud, ``Audio-visual speech enhancement using conditional variational auto-encoders,'' \emph{IEEE/ACM Transactions on Audio, Speech, and Language Processing}, vol.~28, pp. 1788--1800, 2020.

\bibitem{jung_flowavse_2024}
C.~Jung, S.~Lee, J.-H. Kim, and J.~S. Chung, ``{FlowAVSE}: {Efficient} {Audio}-{Visual} {Speech} {Enhancement} with {Conditional} {Flow} {Matching},'' in \emph{Interspeech}, 2024.

\bibitem{mira_2023_lavoce}
R.~Mira, B.~Xu, J.~Donley, A.~Kumar, S.~Petridis, V.~K. Ithapu, and M.~Pantic, ``La-voce: Low-snr audio-visual speech enhancement using neural vocoders,'' in \emph{ICASSP 2023 - 2023 IEEE International Conference on Acoustics, Speech and Signal Processing (ICASSP)}, 2023, pp. 1--5.

\bibitem{chen_rt--voce_2024}
H.~Chen, R.~Mira, S.~Petridis, and M.~Pantic, ``{RT}-{LA}-{VocE}: {Real}-{Time} {Low}-{SNR} {Audio}-{Visual} {Speech} {Enhancement},'' in \emph{Interspeech}, 2024.

\bibitem{gogate_cochleanet_2020}
M.~Gogate, K.~Dashtipour, A.~Adeel, and A.~Hussain, ``{CochleaNet}: {A} robust language-independent audio-visual model for real-time speech enhancement,'' \emph{Information Fusion}, vol.~63, pp. 273--285, 2020.

\bibitem{zhu_real-time_2023}
Z.~Zhu, H.~Yang, M.~Tang, Z.~Yang, S.~E. Eskimez, and H.~Wang, ``Real-{Time} {Audio}-{Visual} {End}-to-{End} {Speech} {Enhancement},'' in \emph{International {Conference} on {Acoustics}, {Speech} \& {Signal} {Processing} ({ICASSP})}, 2023.

\bibitem{inan19_interspeech}
B.~İnan, M.~Cernak, H.~Grabner, H.~P. Tukuljac, R.~C. Pena, and B.~Ricaud, ``Evaluating audiovisual source separation in the context of video conferencing,'' in \emph{Interspeech 2019}, 2019, pp. 4579--4583.

\bibitem{chuang_improved_2022}
S.-Y. Chuang, H.-M. Wang, and Y.~Tsao, ``\BIBforeignlanguage{en}{Improved {Lite} {Audio}-{Visual} {Speech} {Enhancement}},'' \emph{\BIBforeignlanguage{en}{IEEE/ACM Transactions on Audio, Speech, and Language Processing}}, 2022.

\bibitem{li2018effects}
Y.~Li, F.~Wang, Y.~Chen, A.~Cichocki, and T.~Sejnowski, ``The effects of audiovisual inputs on solving the cocktail party problem in the human brain: An fmri study,'' \emph{Cerebral Cortex}, vol.~28, no.~10, pp. 3623--3637, 2018.

\bibitem{adeel2020novel}
A.~Adeel, J.~Ahmad, H.~Larijani, and A.~Hussain, ``A novel real-time, lightweight chaotic-encryption scheme for next-generation audio-visual hearing aids,'' \emph{Cognitive Computation}, vol.~12, no.~3, pp. 589--601, 2020.

\bibitem{hou2018audio}
J.-C. Hou, S.-S. Wang, Y.-H. Lai, Y.~Tsao, H.-W. Chang, and H.-M. Wang, ``Audio-visual speech enhancement using multimodal deep convolutional neural networks,'' \emph{IEEE Transactions on Emerging Topics in Computational Intelligence}, vol.~2, no.~2, pp. 117--128, 2018.

\bibitem{Chung2018VoxCeleb2DS}
\BIBentryALTinterwordspacing
J.~S. Chung, A.~Nagrani, and A.~Zisserman, ``Voxceleb2: Deep speaker recognition,'' in \emph{Interspeech}, 2018. [Online]. Available: \url{https://api.semanticscholar.org/CorpusID:49211906}
\BIBentrySTDinterwordspacing

\bibitem{ma25c_interspeech}
T.~Ma, S.~Yin, L.-C. Yang, and S.~Zhang, ``{Real-Time Audio-Visual Speech Enhancement Using Pre-trained Visual Representations},'' in \emph{{Interspeech 2025}}, 2025, pp. 61--65.

\bibitem{ma_end--end_2021}
P.~Ma, S.~Petridis, and M.~Pantic, ``End-to-end {Audio}-visual {Speech} {Recognition} with {Conformers},'' in \emph{International {Conference} on {Acoustics}, {Speech} \& {Signal} {Processing} ({ICASSP})}, 2021.

\bibitem{ma_visual_2022}
------, ``Visual {Speech} {Recognition} for {Multiple} {Languages} in the {Wild},'' \emph{Nature Machine Intelligence}, vol.~4, no.~11, pp. 930--939, 2022, arXiv:2202.13084 [cs].

\bibitem{erdogan_phase-sensitive_2015}
H.~Erdogan, J.~R. Hershey, S.~Watanabe, and J.~Le~Roux, ``Phase-sensitive and recognition-boosted speech separation using deep recurrent neural networks,'' in \emph{International {Conference} on {Acoustics}, {Speech} \& {Signal} {Processing} ({ICASSP})}, 2015.

\bibitem{wang_compensation_2021}
Z.-Q. Wang, G.~Wichern, and J.~L. Roux, ``On {The} {Compensation} {Between} {Magnitude} and {Phase} in {Speech} {Separation},'' \emph{IEEE Signal Processing Letters}, vol.~28, pp. 2018--2022, 2021, arXiv:2108.05470 [cs].

\end{thebibliography}







\end{document}